\documentstyle[prl,aps,multicol,epsfig]{revtex}
\begin{document}

\title{Relaxation of the field-cooled magnetization of an Ising spin glass}

\author{T. Jonsson, K. Jonason and P. Nordblad}

\address{
Department of Materials Science, Uppsala University, Box 534, SE-751 21
Uppsala, Sweden}

\maketitle

\begin{abstract}
The time and temperature dependence of the field-cooled magnetization
of a three dimensional Ising spin glass, 
Fe$_{0.5}$¥Mn$_{0.5}¥$TiO$_{3}¥$, has been
investigated. The temperature and cooling rate dependence is found to
exhibit memory phenomena that can be related to the memory behavior
of the low frequency ac-susceptibility. The results add some further
understanding on how to model the three dimensional Ising spin glass 
in real space.
\end{abstract}


\begin{multicols}{2}
%\pagebreak
%\onecolumn
\narrowtext

\section{Introduction}
During the last decades the slow relaxation and the attributed non-equilibrium
phenomena of spin-glasses
have been frequently investigated. The experimental tools have usually been
measurements of the relaxation of the  zero-field-cooled magnetization
(ZFC) \cite{ZFC}, thermo remanent magnetization (TRM) \cite{TRM} or
the ac-susceptibility \cite{ac}. The small but significant
relaxation of the field-cooled (FC) magnetization has been less
investigated and
only a few studies of the relaxation behavior of the FC-magnetization
can be found in the literature
\cite{Lundgren85,Djurberg}.
In this paper this conspicuous relaxation is systematically studied for the
Ising
spin-glass Fe$_{0.5}$¥Mn$_{0.5}¥$TiO$_{3}¥$. The results are
discussed and interpreted in terms of a real space phenomenology that
has been developed in the spirit of the droplet model \cite{FisherHuse}. This
way of interpreting the droplet model has recently been used \cite{Jonsson}
to understand the predicted \cite{Random} and observed \cite{Jonsson,Jonason}
memory effects in the relaxation of the ZFC-magnetization and the
ac-susceptibility of three dimensional spin glasses.
\begin{figure}
    \centerline{\epsfig{width=8.5cm,file=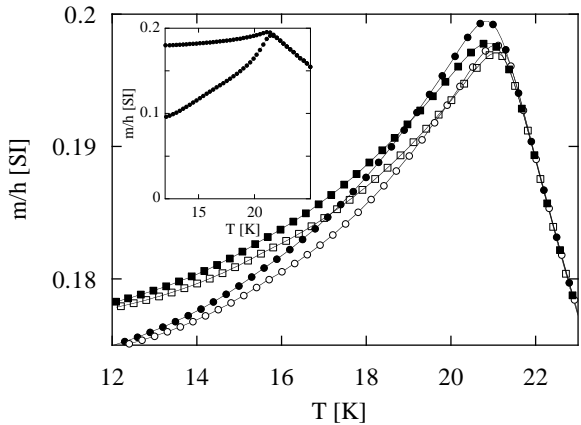}}
    \caption{The magnetization vs. temperature when cooling the sample
    Fe$_{0.5}$¥Mn$_{0.5}¥$TiO$_{3}¥$ in a field, $h=0.5$ Oe,
    to 11.5 K with the rates 0.01 K/min
    (filled squares) and 0.4 K/min (filled circles).
    The subsequent heatings, in the same field, are made with the rate
    0.4 K/min in both cases and are shown with the corresponding open
    symbols. The inset shows zero-field-cooled and field-cooled vs.
    temperature curves in an applied field, $h=0.5$ Oe.}
\end{figure}
\section{Experimental}
The experiments have been performed in a non-commercial dc-SQUID
set-up described in Ref. \onlinecite{Magnusson}.
The magnetic field was induced in a solenoid working in persistent
mode to ensure field stability.
In all experiments the field, $h=0.5$ Oe, was applied at a
temperature well above the spin-glass
transition temperature, $T_{g}¥ \approx 21$ K \cite{Gunnarsson}.

The investigated sample is a
Fe$_{0.5}$¥Mn$_{0.5}¥$TiO$_{3}¥$ ($2 \times 2 \times 5$ mm$^3$) single crystal
which is regarded as a good model system for a
short-range interacting Ising spin-glass \cite{Ito}.

\section{Basic results}
The field-cooled magnetization, $m_{FC}(T)¥$, is measured by cooling
the sample in a constant magnetic field. The field used in this study
(0.5 Oe) is low enough
to yield a linear response in a zero-field-cooled experiment at low
temperatures. In a field-cooled experiment however, any field
causes a non-linearity of $m_{FC}(T)¥$ when passing through the spin
glass temperature $T_{g}¥$. In fact, the magnitude of $m_{FC}(T)¥$ is
governed by the amplitude of the applied field at all temperatures of order
$T_{g}¥$ and
below. In spite of this, to study the dynamics associated with the FC
magnetization we
limit our investigation to using only one but a representative low field.
Fig. 1 shows the FC-magnetization on cooling and heating using
two different cooling and heating rates. The squares represent a
fast rate (0.4 K/min) and the circles a slow rate (0.01 K/min), filled
symbols denote cooling curves
and open symbols the behavior on reheating, at one and the same
heating rate for both curves (0.4 K/min). For reference, the inset
shows the zero-field-cooled and field-cooled magnetization curves for our
sample. As is seen from the main figure, there is a
significant cooling rate dependence of the FC-magnetization. The slow
cooling curve has a higher magnetization than the fast cooling curve at
temperatures just above and through $T_{g}¥$, at a 
\begin{figure}
    \centerline{\epsfig{width=8.5cm,file=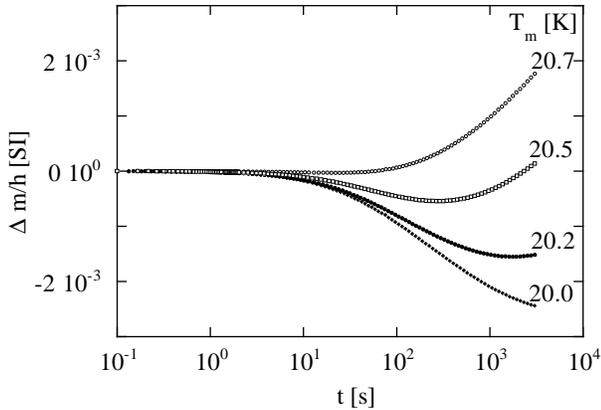}}
    \caption{The relative field-cooled magnetization change  vs. time at some
    temperatures below the transition temperature, $T_{g}¥ \approx
    20.9 K$ for the spin-glass sample Fe$_{0.5}$¥Mn$_{0.5}¥$TiO$_{3}¥$.
    $h=0.5$ Oe.}
\end{figure}
\noindent lower temperature the two curves cross
and the slow cooling curve achieves a lower magnetization value. This
behavior implies a dynamic nature and thus a relaxation of the FC
magnetization if the
sample is kept at constant temperature during cooling. Fig. 2 shows
the relaxation of the FC-magnetization at some constant temperatures
just below $T_{g}¥$ after cooling at a rate of 2 K/min from a temperature well
above $T_{g}¥$. There is an upward relaxation at higher temperatures; a
downward relaxation at short times followed by an upward relaxation at
longer times at a slightly lower temperature; and only a downward
relaxation at even lower temperatures. This behavior is consistent
with the cooling rate dependence observed in Fig. 1. Turning back to
Fig. 1 and looking at the heating curves it is seen that
they always deviate downward from their corresponding cooling curves
until they merge at the equilibrium level well above $T_{g}¥$.
It is however noteworthy that in spite of having the same heating
rate, their temperature dependences still reflect the original
cooling rate all the way up to where they merge at equilibrium, of
certain interest is that also the two heating curves cross at a high
temperature.

Additional information on the non-equilibrium phenomenon of spin
glasses is achieved from intermittent stops at constant temperature
during the cooling process. In Fig. 3 $m_{FC}(T)¥$ at a cooling/heating
rate of 0.4 K/min is drawn. The sample has in these
measurements been intermittently kept at constant temperature for two
or three hours at (a) 18 K, (b) 15 K and (c) both at 18 and 15 K. The result of
the stay at constant temperature is that the magnetization relaxes
downward and on continued cooling after the stop, $m_{FC}(T)¥$ remains on a
lower level
than the reference continuous cooling curve. On reheating, the
heating curve merges with the corresponding reference curve only well above
the temperature where the sample was kept at constant temperature. It
is also worth to note that the heating curve `touches' 
\begin{figure}
    \centerline{\epsfig{width=9.6cm,file=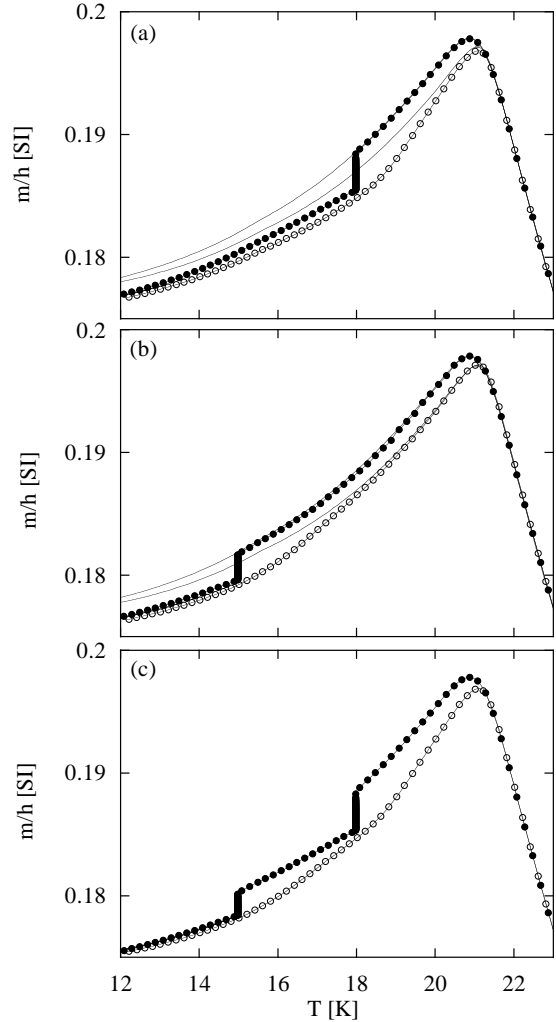}}
    \caption{The magnetization vs. temperature for the sample
    Fe$_{0.5}$¥Mn$_{0.5}¥$TiO$_{3}¥$ cooled in a constant
    field, $h=0.5$ Oe, to 11.5 K (filled circles). The cooling rate of
    0.4 K/min is halted at
    (a) 18 K for 2 hrs, (b) 15 K for 3 hrs, and (c) both at 18 K and 15
    K for 2 hrs and 3 hrs, respectively. The subsequent heatings from
    11.5 K in the same field are shown with open symbols.
    Included for
    reference purposes, shown with solid lines, are the
    cooling and heating curves obtained
    with the rate 0.4 K/min but without intermittent halts. }
\end{figure}
\noindent its cooling curve
just at the temperature of the intermittent stop, not only for one but also
when two or three well separated intermittent stops have been imposed
during cooling. To further illustrate this
behavior we have in Fig. 4 compared the heating curves for the
experiments where the sample has been intermittently kept at constant
temperature
during cooling to the reference behavior without such stops. What
is plotted in the figure is then $\Delta m = m_{FC}(T)¥$-$m_{refFC}(T)¥$. 
Fig. 4a shows
such difference plots for experiments where one intermittent stop has
been made during cooling, circles show the results when the sample was
kept at constant temperature at 18 K for 2 hours, squares \, at \, 
15 \, K 
\, for
3 hours and  
\begin{figure}
    \centerline{\epsfig{width=8.5cm,file=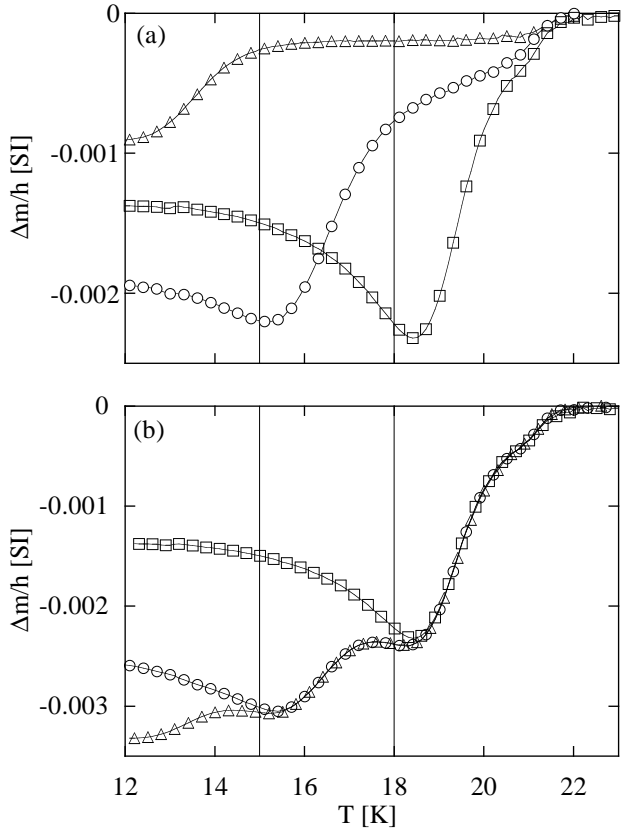}}
    \caption{Relative heating curves 
    $\Delta m(T)/h;$ $\Delta m(T)$ = $m_{FC}¥(T) - m_{FCref}¥(T)$.
    $m_{FCref}¥(T)$ is measured after 
    continous cooling at 0.4 K/min to 12 K and thereafter recording 
    the magnetization on heating at 0.4 K/min. $m_{FC}¥(T)$ is measured 
    on heating the sample at 0.4 K/min, after cooling the sample at 
    0.4 K/min and employing intermittent halts at different 
    temperatures. (a) One halt at 18 K for 2 hrs (squares); one halt 
    at 15 K for 3 hrs (circles); and one halt at 12 K for 3 hrs (triangles). 
    (b) One halt at 18 K for 2 hrs (squares); one halt at 18 K for 2 
    hrs and one halt at 15 K for 3 hrs (circles); and one halt at 
    18 K for 2 hrs, one halt at 15 K for 3 hrs and one halt at 12 K 
    for 3 hrs (triangles). $h=0.5$ Oe.}
\end{figure}
\noindent  triangles at 12 K for 3 hours. Fig. 4b shows the
corresponding curves, but now sequences of intermittent stops at the
same temperatures and times as used in Fig. 4a have
been imposed on the sample during cooling. Squares mark the curve with only one stop at 18 K (the same curve is also drawn in Fig.
4a), circles mark a curve where two stops were imposed, the first at 18 K
and the second at 15 K, and the triangles mark the
behavior when stops were made at 18, 15 K and 12 K. All curves
coalesce at temperatures above their common stop temperatures! It is
remarkable that the downward relaxation accumulated at the low temperatures
does
not sustain at temperatures above the temperatures of the intermittent
stops.
\section{The droplet model}
To interpret the experimental results discussed above we use a
phenomenology based upon
the droplet model developed for Ising spin-glass systems by Fisher and Huse
\cite{FisherHuse}.
The ground state configuration, $\psi$, is in the droplet model two-fold
degenerate by a global spin-flip; $\tilde \psi$. Furthermore,
adopted in the same theory
and very essential for the following discussion of the experimental results,
is the concept of
chaos with temperature \cite{BrayMoore},
i.e. the ground state spin configuration $\psi$ rearranges completely
if the temperature is changed any amount,
$\psi = \psi (T)$. However, there is a length scale,
$l(\Delta T)$, within which no essential change in the ground state
configuration can be observed after
a temperature change, $\Delta T$. This so called overlap length is
trivially, infinite,
at $\Delta T=0$ but decreases rapidly toward zero with increasing $\vert \Delta
T \vert$.
To demonstrate some consequences of this spin-glass model
it is instructive to make a hypothetical
experiment where an Ising spin-glass system is quenched in zero field
from infinite temperature to a temperature $T_{1}¥$ in the spin-glass
phase \cite{Jonsson}.
At $t=0$ after the quench to $T_{1}¥$ there are two ground
states, $\psi (T_{1}¥)$ and $\tilde \psi
(T_{1}¥)$, to map each spin on. In average, half of the spins will
map to $\psi$ and defines $\psi$-domains while the other half of the spins
defines the $\tilde \psi$-domains. This identification forms a fractal domain
pattern which means
that domains of many sizes exist at $t=0$ after the quench.
In a percolation problem
like this, when excluding the infinite domains (clusters), there is a
characteristic maximum size of the fractal domain sizes
which is set by the percolation threshold, $p_{c}¥$ compared
to the site occupation probability, $p=0.5$.
In this simple picture where only temperatures well below $T_{g}¥$
are considered, the domain wall thickness can be
defined and is one lattice spacing.
In order to introduce dynamics into the present, quenched, spin
configuration it is assumed in the droplet model
that collections of ordered spins coherently can flip their
directions anywhere in
the sample. Such a droplet
excitation is associated with an energy barrier, $B$, that must be surmounted.
Fisher and Huse \cite{FisherHuse} made the ansatz that the barrier grows as
\begin{equation}
	B \propto L^\psi
\end{equation}
where $L$ is a measure of the linear size of the droplet and $\psi$ is
a barrier exponent. This is a thermally activated
process implying that larger and larger
energy barriers can be surmounted as time evolves. The linear size,
$L$,
of the largest droplet excitations after the time, $t$, is typically
\begin{equation}
	L \propto \left({\frac{T \ln(t / \tau_{0}¥ ) }{\Delta (T) } }
	\right)^{1 / \psi}
\label{droplet}
\end{equation}	
Here $\Delta (T)$ is an energy scale that increases with decreasing T
and $\tau_{0}¥$ the microscopic spin flip time.
One key point when trying to grasp the dynamics of this spin
glass model is that the quenched spin glass at $t=0$ is a static
system with fractal spin glass domains of many sizes. It is only with
time at constant temperature or age, $t_{a}¥$, that the system can
start to equilibrate through droplet excitation. It is also important
to remember that droplets of a specific size always only include a fractional
part of the total number of spins in the system.
Note that droplet excitations of size $L(t)$, where $t$ is a
time much less than $t_{a}¥$, have reached
equilibrium in the sense that the number of these droplets is
constant.
These droplet excitations can be regarded as paramagnetic
fluctuations when the system is observed at an observation time,
$t_{obs}¥ \ll t_{a}¥$. One crucial effect of the droplet excitations is also to
remove domain structures of size $L(t)$ and smaller, i.e. as the age of
the system increases, a larger and larger part of smaller domain
structures are washed out. Since however, the droplets of a specific
size always are scarce, a substantially longer time is required to
with help of these and larger droplet excitations
remove most of the domain structures of size $R \approx L(T,t)$. On the
other hand,
a logarithmic time perspective must be used when dealing with these slow
dynamics
and the result of this discussion
can roughly be compared to the result by
Fisher and Huse \cite{FisherHuse}, who conclude
that the minimum distance, $R$, between domain walls typically is
\begin{equation}
	R \propto L(T,t)
\label{domain}
\end{equation}
after an equilibration time, $t$.
In summary, the effect of the droplet excitations on the original
quenched domain pattern after an aging time $t_{a}¥$ is: domains of size
$R \ll L(T,t_{a}¥)$ have effectively been
washed out, whereas the number and size of fractal domains of size
$R$ $\approx$$L(T,t_{a}¥)$ and larger
are yet essentially unchanged. The effect of the aging on these larger domain
structures has been frequent but only fractional displacement of their
domain walls. Thus, spin structures on these large length scales
still remain essentially
unaffected by the aging process.

When a magnetic field is applied to a spin glass at a low
temperature, the magnetization process is governed by polarization of
droplets. The composition of a droplet is a collection of spins of
random directions. This implies that the fluctuation in the ratio between
up and down
spins is typically $L^{d/2}$ where $d$ is the spatial dimension of the
spin glass. A droplet thus carries a magnetic
moment of this order of magnitude. In zero field the moments of the
excited droplets are randomly
directed and a macroscopic sample carries no net magnetization (the
original quenched
spin structure that is mapped to the equilibrium configuration is of
course also assumed to have zero net magnetization).
When applying a weak field to a ZFC spin-glass system,
however,  the sample becomes magnetized through a polarization of
droplets that is linear in field at low enough field strengths. Small
droplets are first polarized and as time increases, larger and larger
droplets become polarized and the magnetization continuously increases.
It is always only the very largest active droplets that contribute to the
increase of the
magnetization, the smaller droplets rapidly reach a stationary state
where an equal number of droplets are polarized and depolarized
simultaneously.
As long as the time after the field application (observation time) is
much shorter than the age of the system, only thermalized
paramagnetic droplets contribute to the magnetization process.
However, when the
observation time, $\log$ $t$ becomes of order $\log$ $t_{a}¥$, the largest
droplets
coexist with and also annihilate domain structures of the same size. In
this region the number of excited droplets of this size is larger than the
equilibrium number. This is seen from a larger magnetization value and
an increase of the relaxation
rate of the magnetization at these time scales \cite{ZFC}
compared to the magnetization and rate of the same system at the same
observation time when measured
at quasi-equilibrium (i.e. measured on a system where $t \ll t_{a}¥)$.
(It is worth to notice that this enhanced number of droplet
excitations of size $L(t_{a}¥,T)$ also implies that the domain structures
of this size are more rapidly annihilated than only the
equilibrium number of droplet excitations would allow.) We assume that the
polarization of droplets of any size
always scales with the number of excited such droplets. Consequently,
the equilibrium magnetization, obtained at infinite time, is governed
by `paramagnetic' droplet excitations on all length scales occurring
anywhere in the sample. The droplet picture we advocate here thus
prescribes that the magnetization process is governed by the existing droplet
excitations: on short time scales, the number of droplets is
stationary, whereas on time scales of the order of the age of the
system there is an enhancement of the number of excited droplets
compared to the stationary state. The largest size of active
droplets is governed by the age of the system. An important consequence
of this is that experiments on short observation times only probe
a minor part of existing droplet excitations, whereas essentially all excited
droplets (from the smallest to the largest) are polarized to `equilibrium'
in an experiment where the
observation time and the age of the system are of a similar order of
magnitude. We then observe an ensemble of droplets that is adequately
magnetized with respect to its momentary distribution of droplets and
the applied field. Of
course, this system still experiences a magnetic relaxation since
larger and larger droplets become active and polarized with time. As the
age of the system increases, the enhanced number of droplets appearing
at the momentary age of the system relaxes toward the equilibrium
number, there is thus a
decreasing number of droplets of a specific size with increasing age
of the system. This is
directly observed in a decay of the out-of-phase component of the
low frequency ac-susceptibility when the sample is kept at constant
temperature\cite{Jonason}. The ac-susceptibility probes the system at one
constant
observation time, $t$=$1/\omega$ and the magnitude of the out-of-phase
component gives a direct measure of the
number of droplets of the size $L(T,1/\omega)$. It is
also interesting to note that it can be shown from low frequency
ac-susceptibility
experiment that a large length scale domain structure can be imprinted on
the system by keeping the
sample intermittently at constant temperature at a specific temperature
during cooling, then after cooling to a lower temperature and
reheating the sample, there will be a memory of the
intermittent stop imprinted in the domain structure that is seen as a
dip in the out-of-phase component of the susceptibility at this
temperature \cite{Jonason}.

The basis for understanding the relaxation of the magnetization in a
spin glass lies in the domain structure and how it is affected by the droplet
excitations. These properties are essentially unaffected by a weak
applied magnetic field. Changes of the magnetization are always caused
by polarization or depolarization of droplets of a size corresponding
to the observation time of the experiment. In a FC experiment, the
field is always applied during the experiment and the relevant droplet
excitations that
cause a magnetization change are those of size $L(T,t_{a}¥)$. Two effects
contribute to the relaxation; (i) depending on if the initial polarization
of the droplet excitations (due to the
finite magnetization of the sample) spontaneously is larger or smaller than
the adequate polarization of these, an increase and a decrease of the
magnetization will occur, respectively, and (ii) the changing number of these
droplets as the age of the system increases causes some
relaxation. I.e. there is a
possibility for a relaxation towards higher magnetization and lower
magnetization depending on temperature, cooling rate and the initial
value of the magnetization.

Before ending the discussion on the droplet picture, we complement the
picture with some sentences on the
effect of the applied magnetic field on the equilibrium spin glass phase.
It is contained in the droplet theory
\cite{FisherHuse} and supported by experimental results \cite{Mattsson} that
any applied field field destroys the thermodynamic spin glass phase.
This fact is important for magnetization experiments when passing through
$T_{g}¥$ and also for the
magnitude of the magnetization achieved at low temperatures in a FC
experiment. However, the dynamic response to a small field change in
the spin glass state is linear and the field does not affect the
ongoing equilibration process. The effect of the field is to
limit the maximum size of the droplets on a length scale well beyond
the sizes obtained on our experimental time scales, except very close
to $T=T_{g}¥$ \cite{Mattsson}.
\section{Discussion}
When discussing the
continuous cooling processes of the sample it is convenient to introduce an
effective age of the system, $t_{c}¥$, which is set by the interplay
between reinitialization of the system due to chaos with changing
temperatures, the overlap length and the
equilibration process described by Eq. (\ref{domain})
\cite{Jonsson}.
A slower cooling rate allows the
equilibration process to proceed to longer length scales, which yields a
higher effective age,
$t_{c}¥$. The age of the system defines the
largest size of active droplet excitations and thus also the
typical distance between remaining domain walls $R(T,t_{c}¥)$ at each
temperature, $T$, according to Eq. (\ref{domain}).
In Fig. 1 $m_{FC}¥$ was shown vs. $T$ when cooling the sample
with the rates 0.4 K/min and 0.01 K/min from a temperature well above
$T_{g}¥$. The subsequent heating curves, after
these two cooling procedures, have been obtained with the rate 0.4 K/min
and are shown with the corresponding open symbols.
Regarding the cooling curves,
it was observed that the magnetization corresponding to the slower
cooling rate is larger at high temperatures while smaller at lower
temperatures. When cooling from a high temperature, the correlation
length increases due to critical slowing down and, thus, 
spin glass
correlated regions can form on larger and larger
length scales. When the system falls out of equilibrium at
a temperature above $T_{g}¥$ this means
that droplets with larger relaxation times than the
effective age of the system have not yet been excited. To polarize 
these a slower cooling rate is required. This implies
that the amplitude of the magnetization should increase with decreasing
cooling rate in
this temperature region. On continued cooling through $T_{g}¥$ to
lower temperatures, the correlation length becomes infinite at $T_{g}¥$ and
the system remains critical also below $T_{g}¥$ due to the chaotic nature
of the
spin glass phase. However, the length scale or size of the largest droplet
excitations
is limited by the effective cooling rate and thus this size rapidly
decreases with
decreasing temperature. Considering this fact, the change of the
magnetization of
the sample will depend upon whether the initial spontaneous polarization of
the droplet excitations due to the magnetization of the sample is
larger or smaller than the `equilibrium polarization' for
these droplets. Apparently, from the lower magnetization for the slower
cooling rate at lower temperatures, the spin glass has achieved a too
high magnetization when passing through $T_{g}¥$ compared to the
`equilibrium polarization' for the excited droplets on time scale
$t_{c}¥$. It also worth to note that the heating curves, which are
recorded at one and the same rate (0.4 K/min) do carry a memory of
their different cooling rates. It is especially noteworthy that the
curve corresponding to the slow cooling rate even cross the curve
corresponding to fast cooling rate to
again become more magnetized at a higher temperature. There is a memory of
the cooling procedure
imprinted on the system, we will discuss this memory effect further in
connection
with the experiments employing intermittent stops during cooling.

In Fig. 2 the relaxation of the FC-magnetization at constant
temperature at some temperatures
just below $T_{g}¥$ were shown. The behavior of course accords with the
indications from the cooling rate dependence, in that there is an
upward relaxation at the highest temperature and a downward
relaxation at lower temperatures. The fact that, at intermediate
temperatures, the magnetization first relaxes downward and then at
longer time scales relaxes toward higher magnetization does support
the droplet consequence that the relaxation is governed only by the
largest droplet excitations and that the `equilibrium polarization'
for these can be to large compared to the spontaneous polarization on small
length scales (short times)
and to small compared to the `equilibrium polarization' on large
length scales. Since we do not know where the thermodynamic
equilibrium magnetization for a spin glass is to be found, it is
not possible to, from the measured relaxation of the field-cooled
magnetization, extrapolate the thermodynamic equilibrium value. It is not even
possible to determine
whether this value is larger or smaller than the measured
FC-magnetization at any low temperature in the spin glass state.
It is however possible to predict that at long enough
time scale, there will always be an upward relaxation of the FC
magnetization, since the original mapping prescribes a largest cut-off
length scale for the spin glass domains, whereas droplet excitations
are allowed on all length scales and these largest length scales
droplets will be polarized by an applied field toward a larger
magnetization value. It is possible that the upward relaxation
observed on long time scales at high temperatures just around $T_{g}¥$ is a
consequence
of this prediction.
It should be added that the constant
level of the magnetization seen for the shortest time
scales in Fig. 2 can
be attributed to the effective age of the system. The figure is misleading
in the sense that $t=0$ on time axis actually corresponds
to an age $t_{c}¥ \approx 10$ s which has been obtained during the
finite cooling rate.

Fig. 3 and Fig. 4 showed as discussed above consequences of one or
more intermittent stops during cooling and the influence of these
stops on the subsequent continuous heating curves. The results may be
considered in the light of our droplet phenomenology. During cooling,
the maximum size of excited droplets is determined by the temperature
and the effective age, $t_{c}¥$, of the system. At the constant temperature of
an intermittent stop, droplet excitations are free to occur on length
scales only limited
by the time at constant temperature, $t_{a}¥$. Thus, domain structures
are removed up to the length scale of order of the largest droplet
excitations, $L(T,t_{a}¥)$. These large length scale excitations also cause
a downward relaxation of the magnetization. When the cooling is
recaptured, the size of the droplet excitations are again restricted
by $T$ and $t_{c}¥$. However, droplet excitation of all sizes up to
$L(T,t_{c}¥)$ always occur, these excitations strive to reach the
`equilibrium polarization' adequate for the actual distribution of excited
droplets. The trend, on decreasing the temperature and from subsequent
stops at lower temperatures, is that the
magnetization always decreases with time and temperature. It should
also be emphasized that the decrease of the magnetization obtained
during an intermittent stop is essentially maintained on continued
cooling. The
magnetization does not asymptotically approach the continuous cooling curve
(see Fig. 3). On
reheating, the magnetization falls below the cooling curve,
indicating that some additional downward relaxation occurs on the length
scales of
the cooling process ($t_{c}¥ \approx t_{h}¥$). However, when the
temperature for an
intermittent stop is reached, a high degree of the originally obtained
ground-state order is 
recovered. The relaxation processes that have occurred at lower
visited temperatures just appear as small and dilute regions with a lower 
magnetization in average. Due to the faster dynamics at this 
higher temperature these regions are quickly 
incorporated in the larger domains and hence the magnetization in 
these regions return as well.
Heating through and above the temperature \, of \, the stop
implies that longer 
\begin{figure}
    \centerline{\epsfig{width=8.5cm,file=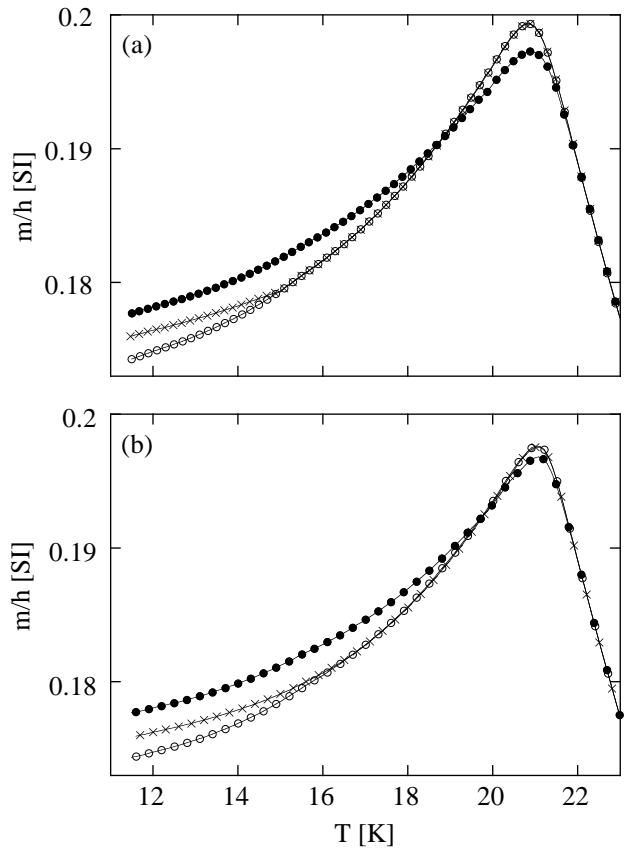}}
    \caption{(a) The solid line and the filled symbols show the
    field-cooled magnetization
    obtained with the cooling rate 0.01 K/min and 0.4 K/min,
    respectively. Open symbols show the magnetization when
    cooling the sample with the rate 0.01 K/min to 15 K followed by the rate
    0.4 K/min from 15 K to 11.5 K. After these different cooling
    procedures the sample is heated with the rate 0.4 K/min and these
    curves are shown
    in (b) with the corresponding notation.
    $h=0.5$ Oe.}
\end{figure}
\noindent length scales are explored and when the temperature has reached
out of the overlap region
around the stop temperature, the heating curve coalesce with the
reference heating curve. It should again be stressed that as Fig. 4b
shows, the FC heating curves all coalesce at high temperatures when
one, two or three intermittent stops have been made during cooling. A
behavior that shows that the spin glass can record the memory of
several well separated temperature stops during cooling in one and the same
ensemble of interacting spins. The key to an understanding of this
ability is the chaotic nature and the different length scales that are
affected by
the largest droplet excitations in the system at the different stop
temperatures.

These memory effects can also be
illustrated by heating curves which are followed by a cooling
procedure where the rate has been changed at some temperature.
Open symbols in Fig. 5 (a) shows the FC-magnetization vs. temperature
during cooling the sample with the rate 0.01 K/min to 15 K where the
cooling rate is changed to 0.4 K/min. Filled symbols and the solid line
show the
magnetization during cooling with the constant rates 0.4 K/min and 0.01
K/min, respectively, down to 11.5 K.
The subsequent heating curves are shown with corresponding
notation in Fig. 5 (b) and are
all recorded with one and the same rate, 0.4 K/min.
Apparently, the heating curve which follows after the change in
the preceding cooling rate at 15 K, merges at the same temperature
with the heating curve obtained after cooling the sample with 0.01 K/min to
11.5 K. Above 15 K, and in accordance with the previous presented results, the
system is unaware of the magnetization history which have occurred at
temperatures below 15 K. The
age, and hence also the magnetization, at each temperature is to a
large extent determined during the preceding cooling process.
\section{Conclusions}
The relaxation of the magnetization of a three dimensional Ising spin
glass can qualitatively be understood in terms of a real space droplet
phenomenology. The response to a field change (e.g in a zero-field-cooled
magnetization or an ac-susceptibility experiment), both the short
time scale equilibrium response and the non-stationary relaxation on
time scales of the order of the age of the system, can adequately be
accounted for only by a polarization of droplet excitations. The
memory behavior is also nicely accounted for in this picture
\cite{Jonsson}. Also the relaxation of the FC-magnetization
reported in this paper can qualitatively be accounted for within this
droplet picture. There are however still important deficiencies as
to quantitative measures in the phenomenology, e.g.
we do not know where the equilibrium magnetization of
the spin glass is to be found and we also do not know the momentary
`equilibrium polarization' of the ensemble of excited droplets at a
specific age of the system. I.e. the direction of the relaxation of
the FC-magnetization is entirely determined empirically and only
indicates the direction toward the momentary equilibrium but it
does not allow an extrapolation to the thermodynamic equilibrium
magnetization. To shed further light on the properties of the
FC-magnetization it would be most useful to employ MC-simulations,
since they allow instantaneous quenches and heatings of the sample and
additionally allow arbitrary and adequately magnetized initial spin
configurations for the relaxation studies.
\acknowledgements
Financial support from the Swedish Natural Science Research Council
(NFR) is acknowledged. The Fe$_{0.5}$¥Mn$_{0.5}¥$TiO$_{3}¥$ single
crystal used in this study has been put to our disposal through a long
lasting and fruitful co-operation with Prof. A. Ito.

\end{multicols}

\end{document}